\documentclass[preprint,12pt,longnamesfirst]{aastex}

\usepackage{latexsym,epsf}

\newcommand \ha   {H$\alpha$}

\newcommand \sii  {[\ion{S}{2}]}
\newcommand \oiii {[\ion{O}{3}]}
\newcommand \rxj  {RX\,J053335$-$6854.9}

\begin{document}

\title{An Intriguing X-ray Arc Surrounding the X-ray Source
\rxj\ toward the Large Magellanic Cloud}

\author{Justin D.\ Lowry\altaffilmark{1}, 
You-Hua Chu\altaffilmark{1},
Mart\'{\i}n A.\ Guerrero\altaffilmark{1,2,3}, 
Robert A. Gruendl\altaffilmark{1},
Steven L.\ Snowden\altaffilmark{4},
R. Chris Smith\altaffilmark{5}}
\altaffiltext{1}{Astronomy Department, University of Illinois, 
        1002 W. Green Street, Urbana, IL 61801;
        jdlowry@astro.uiuc.edu, chu@astro.uiuc.edu, mar@astro.uiuc.edu, 
        gruendl@astro.uiuc.edu}
\altaffiltext{2}{Visiting Astronomer, Cerro Tololo Inter-American 
Observatory, National Optical Astronomy Observatory, which is operated 
by the Association of Universities for Research in Astronomy, Inc.\
under cooperative agreement with the National Science Foundation.}
\altaffiltext{3}{Now at Instituto Astrof\'{\i}sica de Andaluc\'{\i}a (CSIC),
  Spain.}
\altaffiltext{4}{Laboratory for High Energy Astrophysics, Code 662,
 NASA Goddard Space Flight Center, Greenbelt, MD 20771;
 snowden@riva.gsfc.nasa.gov}
\altaffiltext{5}{Cerro Tololo Inter-American Observatory, National 
Optical Astronomy Observatory, Casilla 603, La Serena, Chile; 
csmith@noao.edu}

\begin{abstract}

{\it ROSAT} observations of the Large Magellanic Cloud (LMC) 
have revealed a large diffuse X-ray arc around the point 
source \rxj.
The relative locations of the diffuse and point sources
suggest that they might originate from a common supernova
explosion.
We have analyzed the physical properties of the diffuse
X-ray emission and determined that it is most likely a 
supernova remnant in a low-density medium in the LMC.
We have also analyzed the X-ray and optical observations
of \rxj\ and concluded that it is a foreground dMe star 
in the solar neighborhood. 
Therefore, despite their positional coincidence, these two 
X-ray sources are physically unrelated.

\end{abstract}


\section{Introduction}

The {\it R\"ontgen X-ray Satellite (ROSAT)} observations of the Large
Magellanic Cloud (LMC) have revealed a wealth of diffuse and compact
X-ray sources \citep{Sno94,HP99,SHP02}.
Many large diffuse X-ray sources have sizes greater than 100 pc.
Some of these diffuse sources are associated with superbubbles
and supergiant shells, and others seem to originate from fields
unbounded by interstellar structures identifiable at optical
or radio wavelengths \citep{Dun01,Poi01}.

Among the large diffuse X-ray sources, we have identified two 
objects whose diffuse X-ray emissions have ring morphologies and 
are centered on point sources.
These two objects are intriguing because they are not bounded 
by superbubbles and the relative locations of the point source 
and the diffuse X-ray emission appear to suggest a physical 
association.
The first object, at 5$^{\rm h}$07$^{\rm m}$36$^{\rm s}$,
$-$68$^\circ$47$'$52$''$ (J2000), is projected in the vicinity
of the superbubble N103 surrounding the star cluster NGC\,1850.
A detailed analysis of this object shows that the large X-ray ring,
150 pc in diameter, is most likely a supernova remnant (SNR) formed 
in the low-density halo of the LMC and the central point source 
might be an X-ray binary in the cluster HS122, but the relationship
between the SNR and the X-ray binary is uncertain \citep{Chu00}.

The second object, shown in Figure 1, is projected on the eastern
rim of the supergiant shell LMC-3 \citep{GM78}.
Its $\sim$9\farcm5 angular diameter corresponds to $\sim$140 pc,
if it is in the LMC at a distance of 50 kpc \citep{Feast99}.
The {\it ROSAT} Position Sensitive Proportional Counter (PSPC)
mosaic \citep{Sno94} in Figure 1b shows semi-circular diffuse
X-ray emission centered on the point source RX\,J053335$-$6854.9 
at 5$^{\rm h}$33$^{\rm m}$36$^{\rm s}$,
$-$68$^\circ$54$'$55$''$ (J2000).
The {\it ROSAT} High Resolution Imager (HRI) mosaic \citep{CS98}
in Figure 1c, having a much higher angular resolution, shows 
diffuse X-ray emission in an east-north-west arc centered on 
the point source.
RX\,J053335$-$6854.9 is coincident with a 14th mag stellar object
in the Digitized Sky Survey (DSS).
To determine the origin of the diffuse X-ray emission, we have 
analyzed the {\it ROSAT} observations in conjunction with optical
images and high-dispersion, long-slit spectra of the underlying
10$^4$ K ionized interstellar gas.
To determine the nature of the point source, we have examined 
both the photometric and spectral properties of its optical 
counterpart, and compared them with the spectral  
properties of the X-ray source.
The results of our analysis are reported in this paper.

\section{Observations}

\subsection{{\it ROSAT} X-ray Observations}

{\it ROSAT} observations of \rxj\ and its surrounding diffuse X-ray
emission have been made with both the PSPC and the HRI detectors.
The PSPC is sensitive in the energy range 0.1--2.4 keV and has a
45\% spectral resolution with on-axis angular resolution of $\sim$25$''$
at 1 keV.
The HRI is sensitive at 0.1--2.0 keV; it has a negligible spectral
resolution, but a high on-axis angular resolution of 5$''$.

We have retrieved all {\it ROSAT} observations where \rxj\ falls
within 45$'$ and 20$'$ from the field center for the PSPC and HRI,
respectively.
These observations are listed with their exposure times, locations,
original targets of the observations, and the offset of \rxj\ 
from the field center in Table 1.
Note that for all PSPC observations, \rxj\ lies either close to or
outside the circular window support structure of 40$'$ diameter.
For the former, the wobbling of the telescope moves the source in
and out of the shadow of the window support ring and decreases the
number of counts, while for the latter the detector sensitivity
is reduced and the point spread function is degraded at large off-axis
angles.
Therefore, we used the PSPC observations to carry out only the 
spectral analysis and used the higher resolution HRI observations 
for morphological analysis and comparisons with optical images.

We have used the PROS\footnote{PROS/XRAY Data Analysis System,
http://hea-www.harvard.edu/PROS/pros.thml} software package within
IRAF\footnote{Image Analysis and Reduction Facility, IRAF is
distributed by the National Optical Astronomy Observatories operated
by the Association of Universities for Research in Astronomy, Inc.,
under cooperative agreement with the National Science Foundation.}
for spatial and spectral analyses of these observations.
To improve the S/N, we merged the PSPC observations centered
on SN\,1987A with exposure times $\ge$9 ks
(see Table~1) to produce a 90 ks
equivalent exposure, and extracted spectra from the combined 
data for further analysis.
The other PSPC data were deemed to be less useful and were not used.

\subsection{Optical Images}

Optical images were taken at the Cerro Tololo Inter-American
Observatory (CTIO) with a CCD camera on the Curtis Schmidt telescope.
The observations were part of the Magellanic Cloud Emission-Line Survey
(MCELS) by \citet{Setal99}.
The detector was the SITe2048 \#5 CCD.
Its 24 $\mu$m pixel size corresponds to 2\farcs3.
Images were obtained with the following filters and exposure times:
\ha\ ($\lambda_{c}$ = 6568 \AA, FWHM = 28 \AA), 150 s;
\oiii\ ($\lambda_{c}$ = 5023 \AA, FWHM = 40 \AA), 600 s;
\sii\ ($\lambda_{c}$ = 6738 \AA, FWHM = 50 \AA), 600 s;
green continuum ($\lambda_{c}$ = 5130 \AA, FWHM = 155 \AA), 300 s;
red continuum ($\lambda_{c}$ = 6852 \AA, FWHM = 95 \AA), 300 s.
The \ha, \sii, and red continuum images were obtained on 1998
November 27, and the \oiii\ and green continuum images were obtained
on 2000 December 30.
The emission-line images of \rxj\ and its vicinity are presented in 
Figure 2.

\subsection{High-Dispersion Spectra}

To determine the radial velocity of the optical counterpart of
\rxj\ and to examine the kinematics of the 10$^4$ K ionized gas
underlying the diffuse X-ray emission, we obtained high-dispersion
spectroscopic observations using the echelle spectrograph on the
Blanco 4~m telescope at CTIO on 2002 June 24.
The spectrograph was used with the 79 line~mm$^{-1}$ echelle grating
and the long-focus red camera in the single-order, long-slit mode.
This observing configuration provided a reciprocal dispersion of
3.4 \AA~mm$^{-1}$ and covered the H$\alpha$ and [N~{\sc ii}] 
$\lambda\lambda$6548,6584 lines over a slit length of $\sim$3$'$.
The SITe2048 $\#6$ CCD used has a pixel size of 24 $\mu$m,
which corresponds to $\sim$3.7 km~s$^{-1}$~pixel$^{-1}$ along the
dispersion direction and $0\farcs26$ pixel$^{-1}$ along the slit.
The slit width was $1\farcs8$ and the resultant instrumental
FWHM was 15 km~s$^{-1}$.
The angular resolution, determined by the seeing, was 2$\arcsec$.
The slit was oriented at PA = $345^\circ$ and the total integration
time was 900 s.

\section{Discussion: Physical Nature of the X-ray Sources}

The X-ray point source \rxj\ is projected within the LMC 
and apparently surrounded by the diffuse X-ray arc.
Below we discuss the nature of the diffuse emission and 
the point source individually.

\subsection{Physical Nature of the X-ray Arc}

The circular boundary of the diffuse X-ray emission
surrounding \rxj\ suggests that the hot gas may have
been energized by a supernova explosion at its 
geometric center.
The angular radius varies from 4$'$ on the east side
to 5\farcm5 on the west side.
This diffuse X-ray emission region does not show 
morphological correspondence with optical emission.
As shown in Figure 2, long ($>$ 20$'$) optical filaments 
exist but are associated with the supergiant shell
LMC-3 and do not delineate the X-ray emission region.
From the morphology alone, the hot gas responsible for 
the diffuse X-ray emission does not appear to be 
associated with the cooler gas responsible for the 
optical emission.

It is unlikely that this diffuse X-ray source is 
associated with a SNR in the Galactic plane (within 
500 pc), because its linear size ($<$2 pc) implies a 
very young age and it would have been much brighter.
The probability is also small for this source to originate
from a SNR in the Galactic halo, because massive stars do
not reside in the halo and the low stellar density in the 
halo implies a very low Type Ia supernova rate.
On the other hand, its projected location in the LMC
is within a large-scale star-forming region where the 
supernova rate is expected to be high; therefore, it 
is most likely that a SNR in the LMC is responsible 
for this diffuse X-ray emission.
The linear radius of the X-ray emission region, 60--82 pc,
is larger than those of most SNRs, a few tens of pc in 
diameter, but is reasonable for SNRs formed in a 
low-density medium.

The physical conditions of the X-ray-emitting gas can be 
derived from the observed X-ray spectra, which are a 
convolution of the intrinsic spectra, the intervening 
interstellar absorption, and the PSPC response function.
The contribution of the absorption and response function to the
spectra is energy-dependent, therefore it is necessary to model 
the spectra in order to determine the temperature and emission 
measure of the X-ray sources.
The diffuse emission appears to originate from hot ionized gas
therefore we used the \citet{RS77} models of thin plasma emission
and the \citet{Morr83} models of absorption to simulate the
observed spectra, and determined the best fit by $\chi^2$
minimization.

The best-fit model for the background-subtracted PSPC spectrum 
(see Figure 3a) gives a plasma temperature of
$kT \simeq$ 0.3 keV, an absorption column density
of $N_{\rm H} = 4\times10^{20}$ cm$^{-2}$, and a
normalization factor $A = 10^{10.7}$ cm$^{-5}$.
This plasma temperature is within the range commonly
seen in SNRs.
The rms electron density can be determined from the
normalization factor $A = N_{\rm e}^{2} V/(4\pi D^{2})$,
where $N_{\rm e}$ is the electron density of the plasma, 
$V$ is the volume of the X-ray-emitting plasma, and
$D$ is the distance to the X-ray source.
Assuming a filled hemispherical emitting volume with 
a 70 pc radius, the rms $N_{\rm e}$ is $\sim$0.03 cm$^{-3}$,
which is lower by at least an order of magnitude than most
SNRs in the LMC. 
If the volume filling factor $f$ of the X-ray-emitting 
plasma is less than 1, the density would be $f^{-1/2}$
times higher, but still lower than those of most SNRs.
The total thermal energy of the hot gas is 
$\sim8\times10^{49}$~ergs.
The unabsorbed X-ray flux for the diffuse emission is 
$\sim9\times10^{-13}$ ergs cm$^{-2}$ s$^{-1}$, and the 
X-ray luminosity is $\sim3\times10^{35}$ ergs s$^{-1}$ in the
0.5--2.0 keV band.
Both the thermal energy and X-ray luminosity are similar
to those observed in mature Magellanic Cloud SNRs $\sim10^4$ yr 
of age \citep{Wetal97,Wetal99}.

The physical properties of the diffuse X-ray emission
around \rxj\ are fully consistent with those of a mature
SNR in a low-density medium, much like the large X-ray
ring around RX\,J050736$-$6847.8 \citep{Chu00}.
The low density of the ambient interstellar medium 
explains the large size and the absence of a detectable
optical shell.

The association of the diffuse X-ray arc with a SNR is further
supported by the high-velocity ionized gas detected in our
echelle observation.  
The echellogram in Figure 4 shows two nebular components.  
The first component has a fairly uniform
surface brightness and a nearly constant velocity, at
$V_{\rm Hel}$ = 279$\pm$4 km~s$^{-1}$, along the entire slit
length.
This component originates from a large-scale diffuse
ionized medium in the LMC.
The second component is blue-shifted from the stationary
component, and the velocity offset varies from $<$50 km~s$^{-1}$
at 0\farcm5 south of \rxj\ to $\sim$100 km~s$^{-1}$ at 1\farcm8
north of \rxj.
The high-velocity features are fainter than the large-scale
diffuse component so that only the brightest feature, $\sim20''$ 
south of the star, can be identified with a filament in the 
emission-line images (compare Figures 2 and 4).
The spatial scale and magnitude of velocity offset of this
nebular component are similar to those seen in SNRs in the LMC
\citep{CK88}.
It is possible that this high-velocity nebular component 
is associated with the SNR responsible for the diffuse 
X-ray arc.
While the SNR is in a low-density medium in general, there
may exist small dense clouds which are shocked and give rise 
to the high-velocity nebular emission.
The scarcity of the dense clouds in a low-density medium
prevents the formation of a dense SNR shell structure.

\subsection{Physical Nature of the Point Source}

The X-ray point source \rxj\ appears to be coincident with 
a 14th mag star shown in the DSS.
The position of \rxj\ measured from the HRI image (RH600640N00)
is 5$^{\rm h}$33$^{\rm m}$36\rlap{$^{\rm s}$}{.}0, 
$-$68$^\circ$54$'$55$''$ (J2000).
The star has been cataloged in the Guide Star Catalog 2.2 as 
GSC2.2 S013200256362 at 5$^{\rm h}$33$^{\rm m}$35\rlap{$^{\rm s}$}{.}2, 
$-$68$^\circ$54$'$54$''$ (J2000), in the 2 Micron All Sky Survey
(2MASS) as PSC 05333511$-$6854544 at an almost identical position,
and in the US Naval Observatory-A2.0 catalog as 
USNO-A2.0 0150-03257930 at 
5$^{\rm h}$33$^{\rm m}$35\rlap{$^{\rm s}$}{.}3, 
$-$68$^\circ$54$'$54$''$ (J2000). 
The apparent offset between the X-ray source and the optical star, 
4\farcs5, can be accounted for by the uncertainty in the aspect
solution of the {\it ROSAT} pointing.
Therefore, we consider the X-ray and optical sources coincident.

The photometric measurements of the star given by the catalogs are:
$B = 16.9$ and $R = 14.4$ in USNO-A2.0;
$B = 16.56\pm0.53$ and $R = 14.47\pm0.19$ in GSC 2.2;
and $J = 11.51\pm0.02$, $H = 10.88\pm0.02$, and $K = 10.68\pm0.02$
in 2MASS.
The colors of this star, $B-R = 2.1\pm0.7$, $R-K = 3.8\pm0.2$, and
$J-K = 0.827\pm0.04$, are consistent with a red star with a spectral
type of M2--M3 \citep{Cox00}.
As the distance to this star is unknown, it can be a supergiant in
the LMC, a giant in the Galactic halo, or a dwarf in the solar
neighborhood.

These three possibilities can be distinguished by the radial velocity
($V_{\rm Hel}$) of the star: $\sim$300 km~s$^{-1}$ in the LMC, 
moderate to high velocities in the Galactic halo, and low velocities 
in the solar neighborhood.
We have extracted a sky-subtracted spectrum of this star from our 
echelle observation and plotted it in Figure 4c.
The most prominent spectral feature of the star in this wavelength
range is the \ha\ emission line at  $V_{\rm Hel} \sim 0$ km~s$^{-1}$,
or $V_{\rm LSR} \sim -14$ km~s$^{-1}$.
This small radial velocity strongly argues for this star to be
located in the solar neighborhood.
Thus we further conclude that the star is a dwarf M2--M3 star with
\ha\ emission, one of those commonly called dMe stars.
Adopting a dwarf luminosity class, the distance to the star 
is found to be 60--80 pc.

To determine whether \rxj\ and the star are physically associated,
we examine the X-ray spectral properties.  
The background-subtracted PSPC spectrum of \rxj\ in Figure 3b 
displays a spectral shape typical for stellar coronal emission.
As dMe stars are known to exhibit coronal activity, it is  
likely that \rxj\ and this dMe star are physically associated.
We have used the thin plasma emission model of \citet{RS77}
to fit the PSPC spectrum.  
The best-fit model gives a plasma temperature of $kT = 0.26$ keV
and an absorption column density of $N_{\rm H} = 2\times10^{19}$ 
cm$^{-2}$.
The small absorption column density indicates a small distance.
Adopting the distance of the dMe star, 60--80 pc, the X-ray 
luminosity of \rxj\ is (2--3)$\times10^{28}$ ergs~s$^{-1}$ in 
the 0.5--2.0 keV range.
This X-ray luminosity is completely consistent with those 
expected from dMe stars \citep{Ru84}.
We therefore conclude that the X-ray point source \rxj\
is physically associated with the dMe star.

\section{Conclusions}

{\it ROSAT} observations of the LMC have revealed a large
diffuse X-ray arc in projection around the point source \rxj.
The relative locations of the diffuse and point sources
suggest that they might originate from a common supernova
explosion.
We have analyzed the physical properties of the diffuse
X-ray emission and determined that it is most likely a SNR 
in a predominantly low-density medium.
We have also analyzed the X-ray and optical observations
of \rxj\ and concluded that it is a dMe star in the solar
neighborhood. 
Therefore, these two X-ray sources are physically unrelated.
As dMe stars are the most prevalent stellar X-ray sources in
the solar neighborhood \citep{SFG95}, and as the cooling time
for a SNR in a low-density medium is long, the probability for 
the superposition of a nearby dMe star with a large SNR in the
LMC is not negligible.

\acknowledgements
The project was partially supported by the NASA grant NAG 5-8104.
The optical images from the Magellanic Cloud Emission-line Surveys
were obtained with support from the Dean B. McLaughlin fund at 
the University of Michigan and NSF grant AST-9540747.
This research has made use of the SIMBAD database, operated at 
CDS, Strasbourg, France, and the Digital Sky Survey produced at 
the Space Telescope Science Institute under U.S.\ Government grant 
NAG W-2166. We have also used data products from the 2MASS, 
which is a joint project of the University of Massachusetts and 
the Infrared Processing and Analysis Center/California 
Institute of Technology, funded by NASA and NSF.

\newpage

\begin{figure}

\def\plotonea#1{\centering \leavevmode
\epsfxsize=.50\columnwidth \epsfbox{#1}}

\plotonea{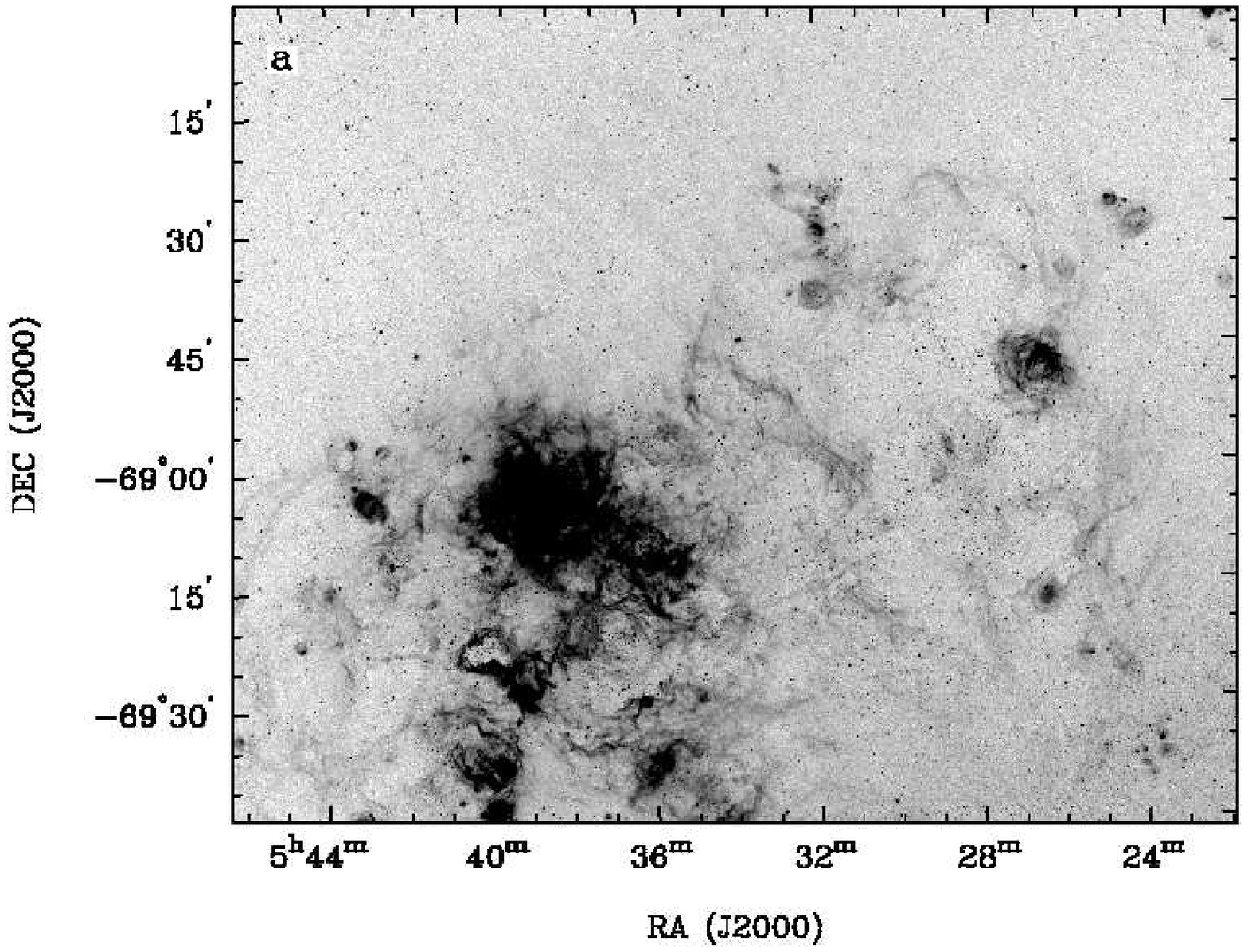}
\plotonea{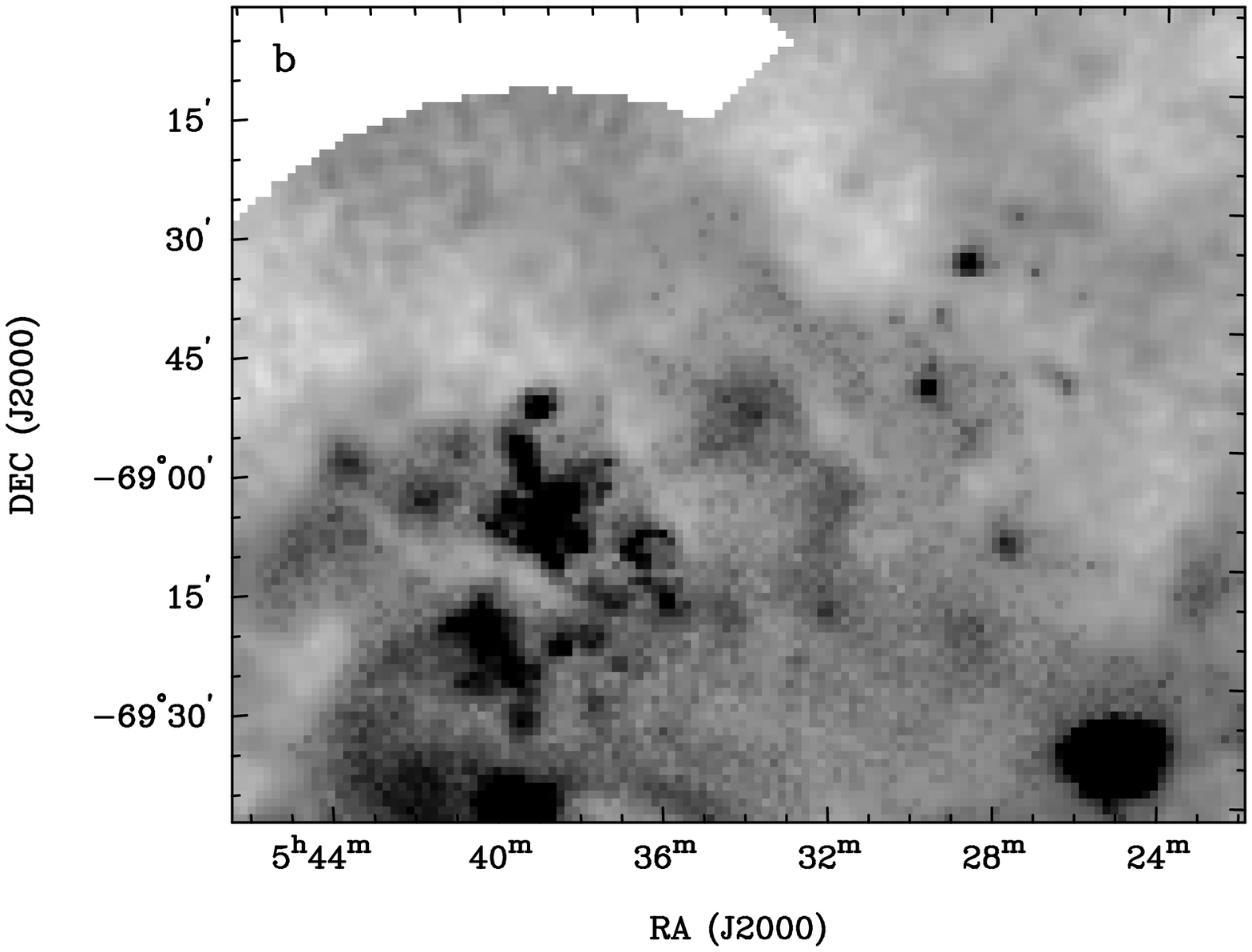}
\plotonea{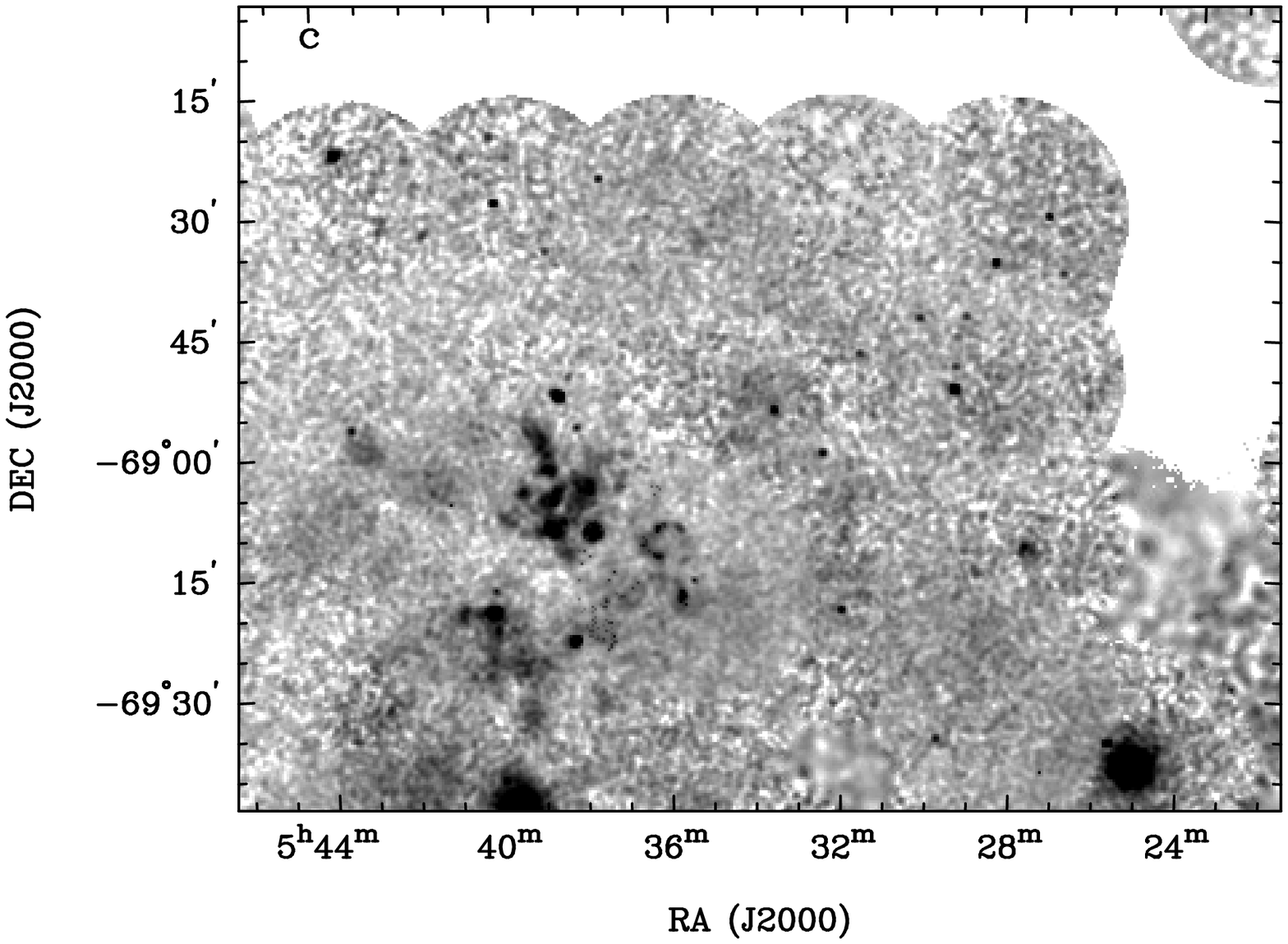}

\caption{(a) H$\alpha$ image of \rxj\ and its vicinity.
(b) {\it ROSAT} PSPC mosaic of the same field in the 0.1--2.4
keV band. (c) {\it ROSAT} HRI mosaic of the same field.
\rxj\ is located at the field center.
}
\end{figure}

\begin{figure}
\plotone{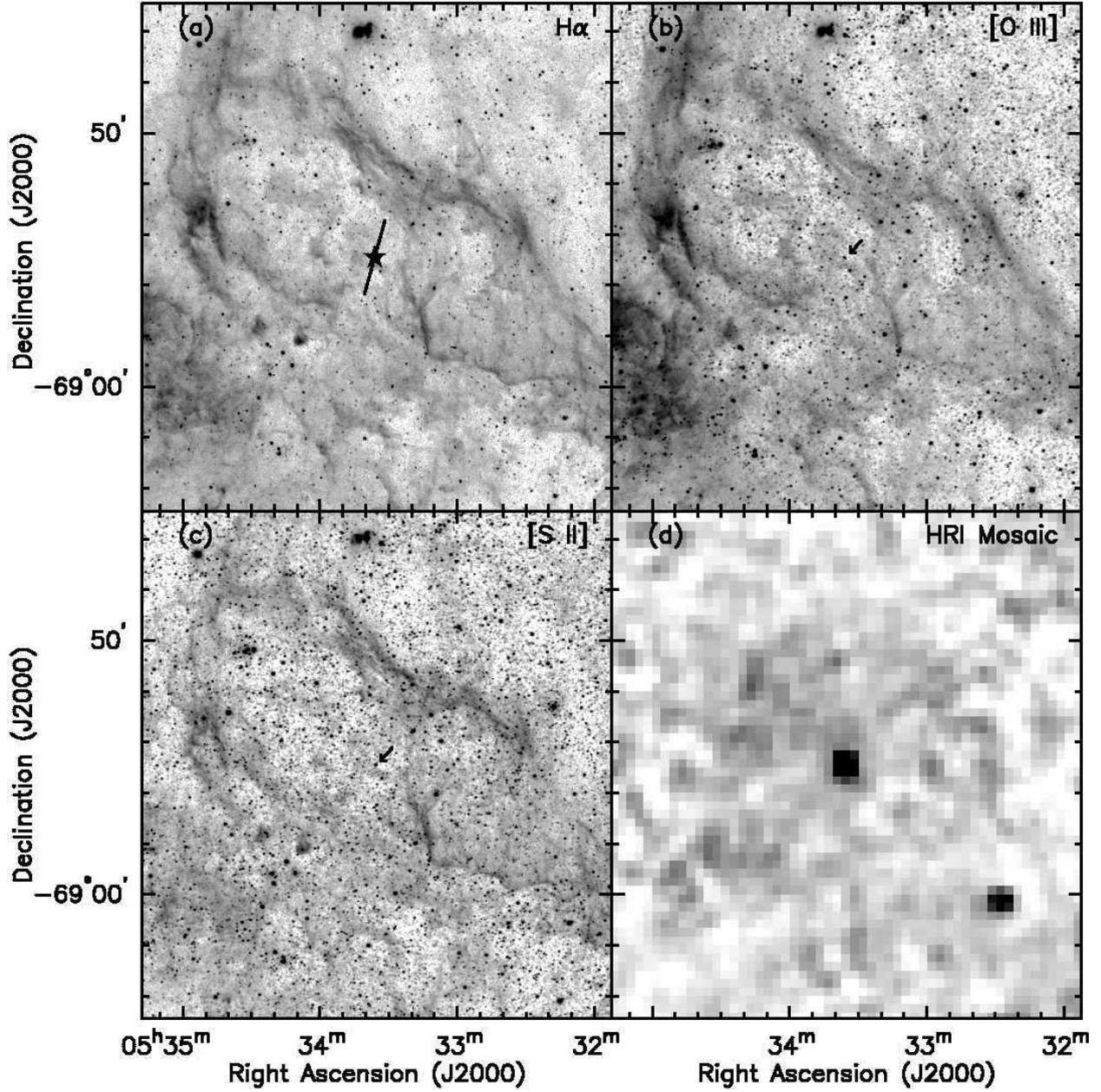}
\caption{(a) H$\alpha$, (b) [\ion{O}{3}], (c) [\ion{S}{2}],
and (d) $ROSAT$ HRI images of the region centered at \rxj.
The stellar counterpart of \rxj\ is marked by ``$\star$'' in (a)
and arrows in (b) and (c).  The slit position of the echelle
observation is marked in (a).
}
\end{figure}

\begin{figure}
\plotone{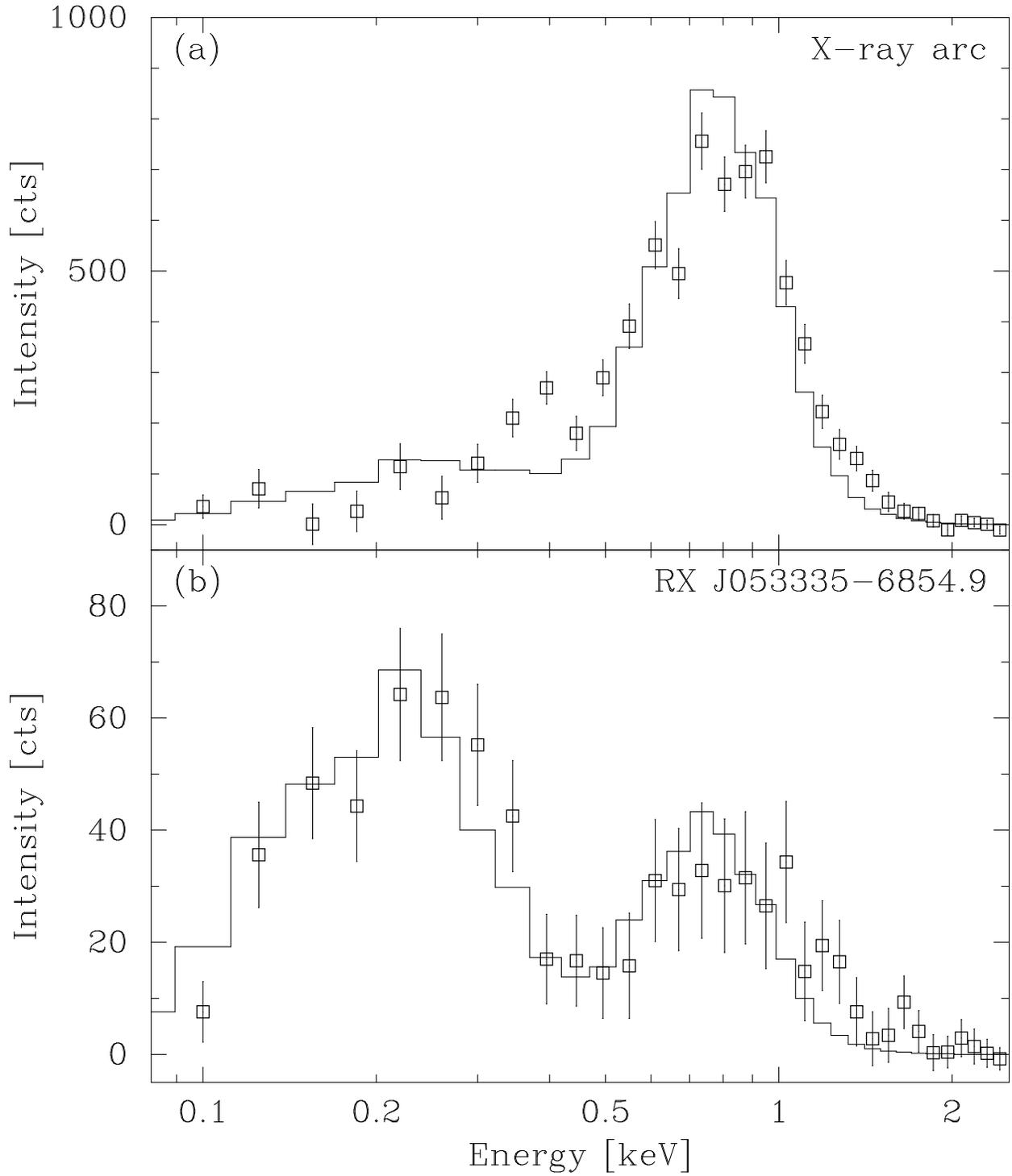}
\caption{{\it ROSAT} PSPC spectra and best model fits
of (a) the diffuse X-ray arc, and (b) the point X-ray
source \rxj.
}
\end{figure}

\begin{figure}
\plotone{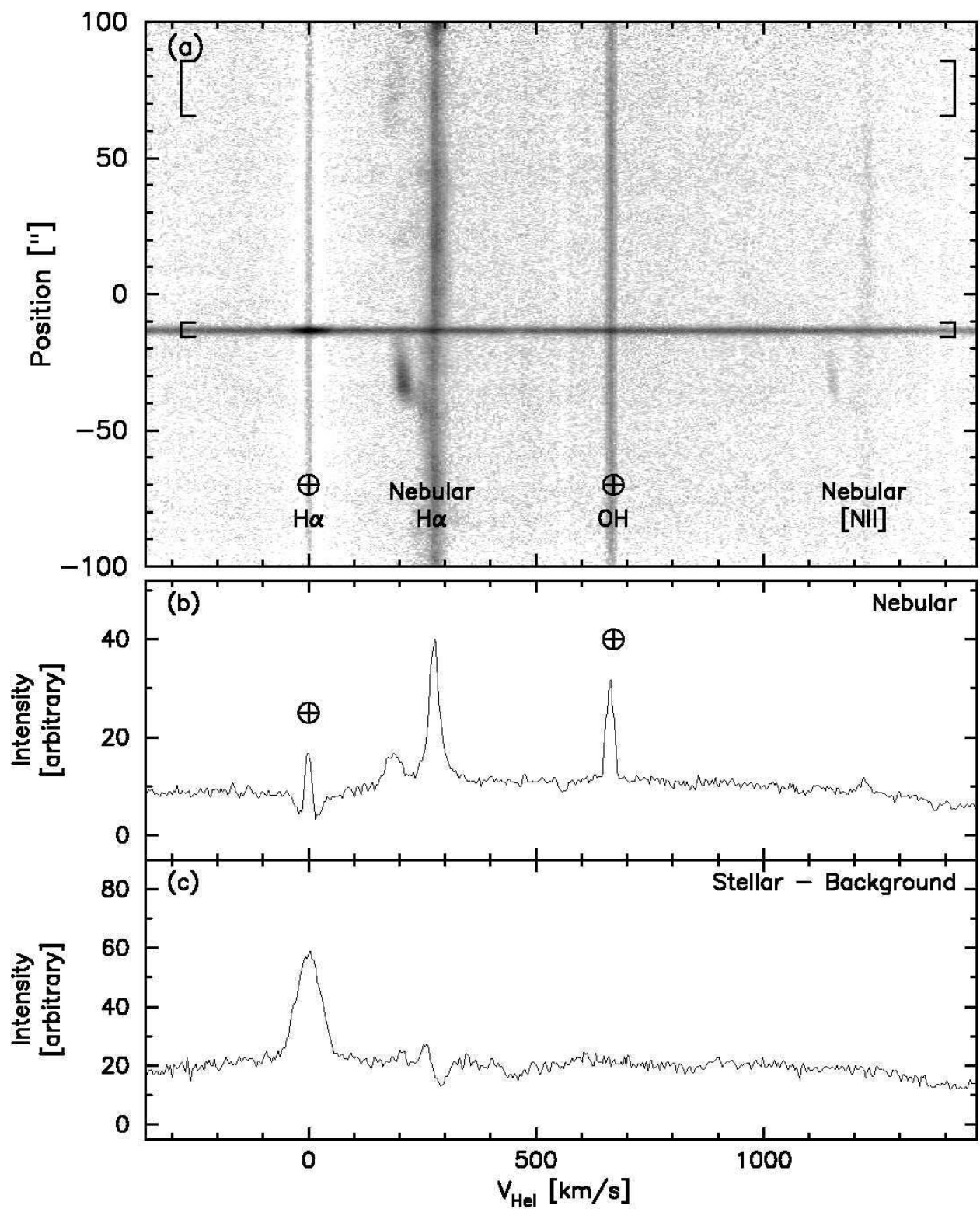}
\caption{Echelle observations of the H$\alpha$+[\ion{N}{2}] 
lines: (a) the echellogram, (b) a representative
H$\alpha$ line profile of the cool ionized gas, and (c) the
H$\alpha$ line profile of the star.  The nebular and stellar
spectra were extracted from the regions marked in (a).
The telluric lines are marked by ``$\oplus$.''  The observation
was made during full moon, so the sky background is dominated
by scattered solar radiation.  The stellar spectrum in (d) is
background-subtracted; the telluric lines are removed, but 
noticeable residuals remain in the nebular H$\alpha$ line.}
\end{figure}

\clearpage

\begin{deluxetable}{lrcclr}
\tablewidth{0pt}
\tablecaption{$ROSAT$ Archive Observations Which Include RX\,J053335$-$6854.9}
\tablehead{ 
Observation & Exposure  & RA (J2000) & 
Dec (J2000) &  Target &  Offset  \\
~~Number\tablenotemark{a} & (s)~~~   & (h ~m ~s~~) & 
(~$^\circ$ ~~$'$ ~~$''$~~) &  Name  &  ($'$) }
\startdata 
RP180179N00     & 15924   & 05 35 28.80 & $-$69 16 11.6 & SN1987A & 23.5 \\
RP180251N00	& 20153	  & 05 35 28.80 & $-$69 16 11.6 & SN1987A & 23.5 \\
RP180294N00	&  2674	  & 05 35 28.80 & $-$69 16 11.6 & SN1987A & 23.5 \\
RP500100A00	& 16957	  & 05 35 28.80 & $-$69 16 11.6 & SN1987A & 23.5 \\
RP500100A01	&  9657	  & 05 35 28.80 & $-$69 16 11.6 & SN1987A & 23.5 \\
RP500140N00	&  2642	  & 05 35 28.80 & $-$69 16 11.6 & SN1987A & 23.5 \\
RP500140A01	& 11625	  & 05 35 28.80 & $-$69 16 11.6 & SN1987A & 23.5 \\
RP500140A02	& 10758	  & 05 35 28.80 & $-$69 16 11.6 & SN1987A & 23.5 \\
RP500303N00	&  9416	  & 05 35 28.80 & $-$69 16 11.6 & SN1987A & 23.5 \\
RP600100A00	&  5803	  & 05 35 38.40 & $-$69 16 11.6 & REGION F& 23.9 \\
RP600100A01	& 16865	  & 05 35 38.40 & $-$69 16 11.6 & REGION F& 23.9 \\
RP500131N00	& 15959	  & 05 38 33.60 & $-$69 06 36.0 & N157 & 29.1 \\
RP500138N00	&  2478	  & 05 26 36.00 & $-$68 50 23.6 & N144 & 38.2 \\
RP500138A01	& 14531	  & 05 26 36.00 & $-$68 50 23.6 & N144 & 38.2 \\
RP500138A02	& 14581	  & 05 26 36.00 & $-$68 50 23.6 & N144 & 38.2 \\
RH600640N00   & 24070  & 05 32 04.80 & $-$68 51 00.0 & LMC POINT 010 & 9.1 \\
RH600634N00   & 23619  & 05 35 52.80 & $-$68 51 00.0 & LMC POINT 004 & 12.9 \\
\enddata
\tablenotetext{a}{RH -- HRI observations; RP -- PSPC
observations.}
\end{deluxetable}

\end{document}